\documentclass[%
 reprint,
superscriptaddress,
 amsmath,amssymb,
 aps,
]{revtex4-2}
\usepackage{hyperref}
\hypersetup{colorlinks=true}
\usepackage{graphicx}
\usepackage{dcolumn}
\usepackage{bm}
\usepackage{mathtools}

\PassOptionsToPackage{normalem}{ulem}
\usepackage{ulem}

\begin{document}

\preprint{APS/123-QED}

\title{A Framework for Evaluating Statistical Models in Physics Education Research}
\author{John M. \surname{Aiken}}
\affiliation {Center for Computing in Science Education \& Department of Physics, University of Oslo, N-0316 Oslo, Norway}

\author{Riccardo De Bin}
\affiliation {Department of Mathematics, University of Oslo, N-0316 Oslo, Norway}

\author{H. J. Lewandowski}%
\affiliation{Department of Physics, University of Colorado Boulder, Boulder, CO 80309, USA}
\affiliation{JILA, National Institute of Standards and Technology and the University of Colorado, Boulder, CO 80309, USA }

\author{Marcos D. \surname{Caballero}}
\affiliation {Center for Computing in Science Education \& Department of Physics, University of Oslo, N-0316 Oslo, Norway}
\affiliation {Department of Physics and Astronomy; Department of Computational Mathematics, Sciences, and Engineering; and CREATE for STEM Institute, Michigan State University, East Lansing, Michigan 48824}

\date{\today}

\begin{abstract}
Across the field of education research there has been an increased focus on the development, critique, and evaluation of statistical methods and data usage due to recently created, very large data sets and machine learning techniques. In physics education research (PER), this increased focus has recently been shown through the 2019 Physical Review PER Focused Collection examining quantitative methods in PER. Quantitative PER has provided strong arguments for reforming courses by including interactive engagement, demonstrated that students often move away from scientist-like views due to science education, and has injected robust assessment into the physics classroom via concept inventories. The work presented here examines the impact that machine learning may have on physics education research, presents a framework for the entire process including data management, model evaluation, and results communication, and demonstrates the utility of this framework through the analysis of two types of survey data.

\end{abstract}

\maketitle

\section{Introduction}

Influenced by the development of machine learning, the statistics community has had discussions on the importance of a model's predictive capability, in addition, or in contrast, to its explanatory power \cite{breiman2001statistical, hastie2005elements, shmueli2010explain}. This argument has been extended to the often used statistical inference models in social science \cite{Hofman486}. This conversation has moved into the Physics Education Research (PER) community through a recent critical examination of quantitative methods used in PER \cite{FocusedCollection}, which included calls to modernize statistical modeling \cite{van2019modernizing}. The framework presented in this paper builds upon that work to produce a method for evaluating statistical studies in PER that is inline with calls for updating current practices in social science fields, such as computational social science and psychology \cite{poldrack2019establishment}.

Traditionally PER has used linear models such as ordinary least squares regression, mixed effects regressions, and logistic regression  (e.g., \cite{gettingStartedPER, pollock2007reducing, lin2017exploring}). Linear models assume that direct combinations of features represents a direct observation \cite{raudenbush2002hierarchical}. These models focus on determining unbiased estimates of model coefficients in comparison to predictive performance. In \citet{shmueli2010explain}'s words, traditional PER models focus on explanatory power rather than predictive power. Models with poor predictive power, as measured by goodness of fit tests, are sometimes still assumed to be transferable and the results generalizable if model parameters are considered statistically significant (e.g., demonstrating via \textit{p}-values and effect sizes). ``Hidden'' effects are are used to explain why there is low explained variance in many models (e.g., \cite{coletta2005interpreting}). Using models in this way is not uncommon in other sciences. However, there are calls within physics and other fields to update model assessment and evaluation to include deeper measures of prediction (e.g., \cite{carleo2019machine, poldrack2019establishment}).

Much of the recent work in the statistics and machine learning communities has focused on prediction rather than explanation. In the former, community data collected is put into a model that is evaluated on its predicted output. The internal structure of the model is not assumed to be representative of the system being studied \cite{breiman2001statistical}, that is, there is not necessarily a correspondence between model structure and system structure and dynamics. The statistics community still considers the model central and highly values its interpretability. Model performance is evaluated using an independent test set of data \cite{hastie2005elements}. These data are independent; they are completely separated from the data used for model building \cite{geron2017hands}. Thus, the models are trained using a ``training data set'' and then evaluated using a ``testing data set.'' Models with poor predictive power are rejected in favor of more predictive models \cite{geron2017hands}. By evaluating the predictiveness, instead of the statistical significance, a relationship between two variables can emerge beyond a ``statistical association'' \cite{poldrack2019establishment}.

The framework that is presented in this paper is designed to provide a method for establishing generalizable results from educational data mining studies \cite{ding2019theoretical}. Its goal is to aid in exploring research questions such as: (1) who will change majors in physics to other majors \cite{aiken2019modeling}, (2) which students in physics courses need extra help to succeed \cite{stewartML2020}, and (3) what patterns exist in student preparation that may impact learning at the university level \cite{prep2020}. Producing generalizable results through use of statistical model evaluation will provide an additional tool to the PER tool box. This tool will allow researchers to compare results across similar university environments.

It is important to note that this framework is not designed to test causal models \cite{ding2019theoretical}, nor is it designed for evaluating latent variable modeling, such as item response theory. While this framework is a good starting point for evaluating these kinds of models too, causal models require more strict verification that is grounded in the theory being tested and cannot rely on solely statistical model evaluations. While this is true in educational data mining studies as well, in an educational data mining study, theory only suggests what variables could be used in a model \cite{ding2019theoretical, pardos}.

 In this paper, we review how statistical modeling has been done in PER and then describe the current discussions in the statistics and machine learning communities on model representation and evaluation. We then provide evidence that the current quantitative PER work could be advanced by using robust model assessment inline with other quantitative fields. Then, we offer a framework for the PER community that considers data management, model evaluation, and results communication holistically. This framework is grounded in work presented in the recently published PRPER focused collection \cite{FocusedCollection}, as well as in modern views from the statistics and machine learning communities. The framework we present is model agnostic, that is, it will work for traditional linear models, as well as more complex ``machine learning'' models. Finally, we present two applications (``case studies'') of the framework: (1) using the Learning About STEM Student Outcomes (LASSO) data set \cite{nissen2018participation}, and (2) using the data from the Colorado Learning Attitudes about Science Survey for Experimental Physics (E-CLASS). These two case studies demonstrate the need for assessing the predictive ability of models beyond what is typically done in quantitative PER.

\section{Background}
\label{sec:background}

Physics education research (PER) is an interdisciplinary sub-field of physics that investigates many aspects of physics education \cite{national2013adapting}. PER uses a wide variety of studies and types of research, including theory building, empirical investigations, qualitative research, and statistical modeling. Quantitative PER has historically focused on investigating how instruction impacts student outcomes such as conceptual learning \cite{hake1998interactive} and attitudinal shifts \cite{wilcox2016students}.  Papers in PER that use statistical modeling typically explore data using logistic or linear regressions. Typically, they do not use Cox style regressions (i.e., time-to-event models) or machine learning models (with few exceptions, e.g., \cite{young2018using, aiken2019modeling, stewartML2020}), although use of these methods is increasing. Early work in PER also generated its own statistics based on empirical observations such as the Hake normalized gain \cite{hake1998interactive}.

\subsection{To Explain or Predict?}
A common goal across all science is the understanding and explanation of how a system works. For example, physics has historically had a close connection between the explanation of how a system works and a prediction of the time evolution of the system. Newton's second law provides an explanation of what causes an object to move, as well as excellent predictions as to where an object will end up after some time $t$. In education research,  we rarely have access to such causal relationships that produce both excellent explanations of a system, as well as excellent predictions. However, it is sometimes assumed that a model that produces excellent explanations is also predictive \cite{shmueli2010explain}. This can conflate the goals of explanation and prediction, which are different aims. This is seen across both the culture of how models are viewed and the practices and choices of the data used for modeling. An explanatory model might rely on a specifically designed survey to, for example, assess a particular attitude towards experimental physics \cite{wilcox2016students} with data from particular types of courses. Whereas a predictive early warning model of student failure in a physics course \cite{stewartML2020} would not be useful without current data from ongoing courses that are actively streamed into the model.

\citet{breiman2001statistical} identified two separate cultures that use statistical models: one culture is grounded in the theory that the  stochastic models that generate the data must be approximated, the other culture is grounded in the theory that the model that generates the data should remain unknown. \citet{breiman2001statistical} argues that the data generating process is normally complex and statistical models, such as ordinary least squares regression or multi-level regressions, cannot approximate it. In  typical machine-learning reasoning, he supports algorithmic methods, in the form of a black-box, which might produce better and different understanding of the data. In this context, a black-box model is one where the way the model arrives at the solution is unknown. Taking the concept to the extreme, he goes on to state: ``the goal [of a statistical model] is not interpretability, but accurate information'' arguing that by sacrificing accuracy for interpretability, a researcher sacrifices understanding between the dependent and independent variables.'' While the advantages of an interpretable statistical model over a black-box solution are evident, in terms of transparency, portability and generalizability, Breiman's statements should be taken into consideration when highlighting the importance of a model's predictive ability. We argue that a black-box that does not give any further insight into the problem is not terribly useful.

Focusing on prediction can lead to the discovery of novel causal relationships, the generation of new hypothesis, and the refining of existing explanatory models \cite{shmueli2010explain}.
Much of recent literature across social science and statistics has called for a change in the types of models used to predict social systems, the theory and frameworks that motivates the choice of models, and the evaluation methods used to demonstrate prediction. \citet{Hofman486} recommends that ``current practices for evaluating predictions must be better standardized,'' arguing that current methods in social science focus too much on explanation and too little on prediction. \citet{Hofman486} recognizes that there is a fear that complex models may lose interpretability, but points to innovations in recent literature that overcome a loss of interpretability in complex models (e.g.,\cite{ribeiro2016should}). The work in PER has not yet fully engaged with this discussion, but work is beginning to discuss the importance of different considerations in conducting quantitative PER \cite{ding2019theoretical}.

\subsection{Quantitative Analysis in PER}

Quantitative methods in PER have evolved over time. There is a robust and growing community of researchers in physics education that are thinking about the entire process of quantitative research from data collection and using more sophisticated models to evaluating those models and presenting results. In this section, we discuss a few salient components of the current state of quantitative PER methods and where improvements might be made. 

\subsubsection{Data Collection}
The PER community has investigated how data preparation and organization impacts observational and experimental results. Both \citet{ding2012getting} and \citet{springuel2019} recognize that education data typically are not gathered as interval or ratio data (as is typical in a traditional physics study). Thus, the encoding of data can directly impact the results reported. \citet{springuel2019} identifies several types of data that are common in PER and describes the limitations of using each type of data in research. These types include asymmetric and symmetrical nominal data, ordinal data, interval data, and ratio data. These types are presented on a scale as to how informative they are and what they are able to do. For example, ordinal data contains a finite number of ordered responses (e.g., a Likert scale). In contrast, ratio data creates an scale of data from 0 to 1 that represents the presence or absence of some observation.

Missing data can also be an issue in quantitative PER studies \cite{nissen2019missing}. Most statistical models are not capable of handling missing data directly. Data must either be removed or imputed. \citet{RubinMissing} describes three types of missing data: (1) missing completely at random, (2) missing at random, and (3) missing not at random. Data missing completely at random is when data are missing and there are no mechanisms that can be demonstrated that explain how the data came to be missing. The other kinds of missing data assume that there is some mechanism that may cause the data to be missing (e.g., low performing students are less likely to complete both a pre and post concept inventory). When these data are missing in low amounts, it is reasonable to do row wise deletion of data that is missing \cite{springuel2019}. Row wise deletion means that data are removed from the data set for an entire row if any of the columns in a row have missing data.

\citet{nissen2019missing} recommended the use of multiple imputation \cite{little2019statistical} to fill in the missing data in cases with larger amounts than discussed by \citet{springuel2019}. \citet{josse2019consistency} demonstrates that in addition to using various imputation methods (including multiple imputation), adding a covariate column that explicitly includes whether data are imputed or not can increase model performance without biasing data, especially when missing data are related to the target variable.

Finally, communication of data context has also been observed to be an issue in PER. \citet{knaubaikending2019} found that approximately half of papers published in Physical Review: PER do not describe basic descriptive information about the students being studied such as their demographics, average grades, or institutional descriptors \cite{demo2020}.

\subsubsection{Statistical Modeling}
Recent efforts have been made to modernize statistical modeling in PER both in methods and in theory. For example, \citet{aiken2019modeling} recommend using out-of-sample data to evaluate classification models. Out-of-sample data are data that are randomly selected from the data set under study that is not used to train the model. Once the model is trained, it is then used to assess model performance (e.g., using mean squared error). \citet{van2019modernizing} recognize that much of the data in social science and in PER can have a hierarchical structure that is not picked up by typical linear regressions with only fixed effects. They recommend using multi-level models to calculate the contributions of the hierarchical structure that include random effects. \citet{theobald2019beyond} make similar recommendations extending the types of regressions that should be considered in PER (e.g., proportional odds, multinomial, etc.) and provide a toolkit to determine the most effective generalized linear model that a researcher should use to analyse their data. This is inline with much of the work of Bryk and Raudenbush in the 1980s culminating in their textbook \cite{raudenbush2002hierarchical}. \citet{bryk1989toward} identified that ``the real need [in education research] is for statistical models that provide explicit representation of the multiple organizational levels typically encountered in educational research.'' \citet{bryk1989toward} identified that learning happens across organizational levels such as an individual student, the classroom, the institution, and even the different times a measurement is taken.

\subsection{The role of theory and domain expertise in statistical model evaluation}
The goal of this paper is to present a framework for evaluating statistical models in PER to allow for better generalizability of results. However, we are not suggesting that the only reliable model evaluation comes from whether the statistical model fits the data input into the model or not. It is also important that the data input decisions, and the results that are interpreted, be done in the context of educational theory. Integrating theory and domain knowledge is currently an open topic in the fields of statistics and machine learning. Within PER, domain knowledge and theory can include, as an example, theories of equity and inclusion \cite{brewequity}, theories of learning, such as the role of interaction \cite{hake1998interactive}, impact of social interaction and social network \cite{brewesocial}, neurological and cognitive theories \cite{redish2004theoretical},
and theories of community integration, such as the communities of practice theory \cite{wenger2000communities}. Statistical modeling allows for a process of rigorous testing of these theories. In reverse, theory allows for strong explanations about when a model fails (e.g., a lack of social network data can impact the overall predictability when examining university student drop out \cite{aikenplosone}).

\begin{figure*}[t]
    \centering

    \includegraphics[scale=0.6, trim=0 1cm 0 0cm]{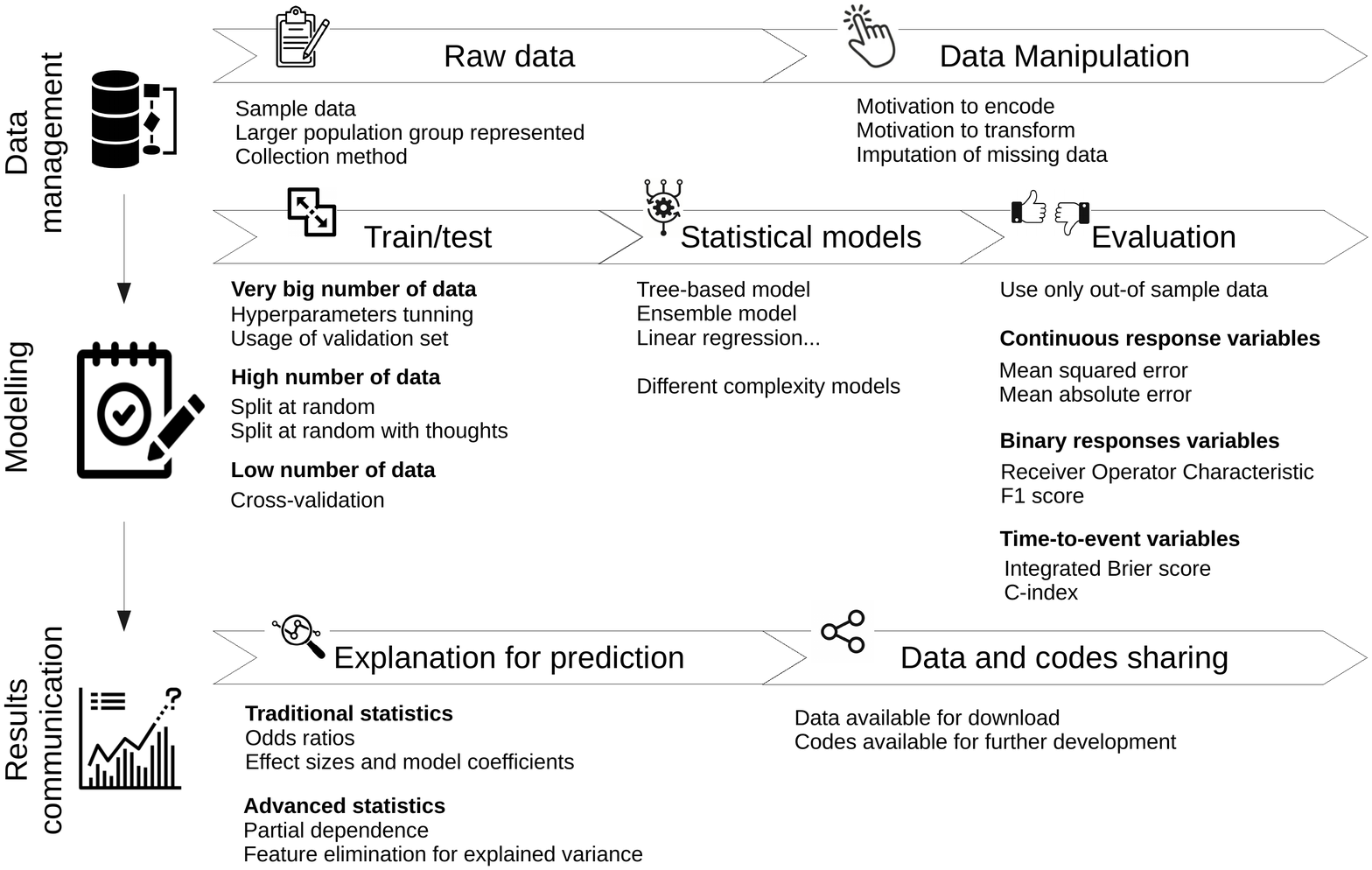}
    \caption{A framework for building and evaluating statistical models in physics education research.}
    \label{fig:framework}
\end{figure*}

\section{A Framework for Evaluating Statistical Models in PER}

The framework presented in this section is aligned with current calls in social science to modernize statistical model evaluation, so that predictive models presented are more robust. Fig. \ref{fig:framework} is a graphical representation of this framework. Each step in Fig. \ref{fig:framework} is discussed below. The framework is shown as a flow diagram to represent each integral step in building a statistical model. This flow diagram  represents explicitly each step in the statistical modeling process. It is not meant to represent an ordered process of steps that are performed in succession until completion. In some cases, it is useful to iterate this process to produce the best models. However, care should be taken to not iterate until results are found (i.e., ``p-hacking'' \cite{phack}). However, this does not mean that successive tests of different theories and statistical models cannot inform the statistical modeling process and ultimately the results of a research study. Additionally, this process can be transparent through sharing of analysis in appendix, code repositories, and other supplemental materials.

Prior to the collection of any data, researchers must identify their theoretical constraints and research questions. These dictate what data is to be collected at all. For example, if researchers are interested in student learning they might consider a cognitive model of learning and thus they would collect data that would indicate that (e.g., aptitude scores, scores on other cognitive measures, prior course work, amount of sleep, etc.). That is, measurements that all effect cognition. Instead if researchers are interested in identity formation and if someone adopts an identity or not, they could collect other data such as time spent on activities related to that identity, a students rating of their own competency, etc. Both are welcome in the analytical framework presented in this paper and underlie all of the data and analysis.

\subsection{Raw data}

Raw data (see Fig. \ref{fig:framework}) collected in educational studies should be described in adequate detail. This includes discussing the sample data that is analysed in the paper and the larger population group that the sample is drawn from. It also means discussing how the data was collected (e.g., in-class, online, for credit, optional, etc.). We recommend following the questions presented in \citet{knaubaikending2019} such as ``what information regarding the sample is useful for the audience,'' as well as emphasizing explicit sample descriptions and limitations \cite{demo2020}.

\subsection{Data encoding, transformations, missing data}

Data collected for any research study is dictated by the research study design and ultimately the research questions being asked. How these data are encoded can impact the results of a model. For example, a student variable could be ``passed course'' or it could be the student's grade. Encoding and transformation procedures applied to data should be described in adequate detail (see Fig. \ref{fig:framework}). We do not recommend to state explicitly what each data type is for various kinds of data sets. Instead, we recommend authors motivate the methods used to transform or encode each variable in a model from the type of underlying data. For example, if data are to be transformed using summary data, such as with a z-score \cite{aiken2016methods}, then the summary statistics (e.g., mean and standard deviation for a z-score) should be calculated only for data labeled as training data. These summary statistics can then be used when calculating the z-score for the test statistic. Otherwise, data leakage could occur, inflating the model's prediction ability \cite{poldrack2019establishment}. Data leakage is what happens when data or information from the test set are included in the training data set. When this occurs, the statistical model can overfit on patterns in the test data set and present artificially higher predictive results. Additionally, in this step, missing data should be addressed. If applicable, we recommend using an imputation method of the authors' choice (e.g., multiple imputation \cite{buuren2010mice} or K-nearest neighbors \cite{crookston2008yaimpute}) and including a missingness indicator for each column with missing data that has been imputed \cite{josse2019consistency}. In some cases, there may be too much missing data and therefore imputing the data may bias the results \cite{jakobsen2017and}. There is no universal solution to missing data. For this reason as well, it is extremely important to report the treatment of missing data in the analysis.

\subsection{Train/test split}
\label{sec:traintestsplit}
This process separates data randomly into training and testing data (see Fig. \ref{fig:framework}). The training data are used to ``teach'' the statistical model the patterns in the data. The testing data are used to measure how 
well the model performs, and should be completely independent from the training data to avoid overestimation of the performance. 
It is important in this step to acknowledge that some data might be stratified in ways that are inappropriate to completely separate data at random (e.g., randomly picking data across time windows in a study that is attempting to predict time to some event). For example, \citet{aiken2019modeling} present a statistical model that predicts if a student enrolled in the physics department will graduate with a physics degree or switch to an engineering degree. In that study, students' high school GPAs had risen over time. One way to adjust model training could be to randomly split data per year instead of randomly across all data regardless of the enrollment year.

When the number of observations is too low to justify a split into a training set and a testing set, more elaborate methodologies can be implemented, for example cross-validation \cite{shao1993linear}. Cross validation splits the data into several (`K') subsets of approximately same size: each subset is in turn used as a test set, in which the performances of multiple models trained on the remaining K-1 subsets. The K performance measures obtained are then averaged to obtain a final evaluation of the model. This process allows for the stability of the model to be tested and permit one to use each data point both in the training and test processes.

In addition to data splitting methods for model evaluation, many modern models have hyperparameters that need to be ``tuned'' to produce a best fit model. Hyperparameters are parameters that describe the structure of the model as opposed to the structure of the data. This could include the depth of a decision tree or the value of a regularization term. When the model fitting procedure requires fixing the value of hyperparameters, such as the number of iterations and the step size for boosting (see Section \ref{boosting} for more details), these must be tuned by using a third set, called validation set. In this case, data are split into three groups: training, validation, and testing. The models are trained using the training set for each value of the hyperparameters, the best values are selected using the validation set, and finally the performance of the model is evaluated on the test set. Even if there is no need to tune an hyperparameter, this three-fold split can be used to improve the quality of the analysis (model selection and model fitting are performed on independent observations), and therefore a stronger claim of generalizability. Using a validation data set typically requires having a very large data set. The split training/validation is mimicked by applying cross-validation on the larger set. Cross-validation is the same train/test separation process, but abstracted to $K$ separate data sets known as ``folds''. Then each fold is compared to the other folds to produce average estimates. When cross-validation is used instead of the training/test set, a further cross-validation procedure is implemented within the K-1 folds used in turn as the training set.

\subsection{Statistical Models}

Here, we emphasize that models is plural. The best statistical model might be tree based (e.g., random forests \cite{breiman2001random} or gradient boosted trees \cite{friedman2001greedy}), or it might be an ensemble of models \cite{hastie2005elements}, or it might even be a model more familiar to PER researchers, such as a linear regression. The goal is to test models of varying complexity and pick a model that has the highest predictive ability with the lowest complexity both in terms of number of features used and how the model itself works. This is already being done in PER in some studies, typically by comparing models with few features to models with many features and reporting in-sample goodness-of-fit tests. For example: \citet{prep2020} compared different models that predict final exam scores to assess the impact of preparation; \citet{aycock2019sexual} compares models that predict a student's sense of belonging in physics due to different experiences. Additionally, more sophisticated models can increase our understanding even when we have lower amounts of data on student sub-populations. For example, using a data set that had a small racial and ethnic minority population (~15\%) \citet{aikenplosone} presented gradient boosted logistic regression models that performed much better than traditionally solved maximum likelihood models when predicting undergraduate student time to graduation. Statistical model choice can impact not only the overall predictability of a model, but also our understanding of a system even when our data on all populations are not equal. It is important to note that the choice of the models to investigate must be done before the data analysis step, but rather in the design of the experiment. Trying all possible methods until one ``fits'' the data are a way of overfitting (looking for significance) and should be avoided. 

\subsection{Evaluate model performance}

A model's predictive ability should be evaluated by its ability to predict new data that the model has not seen before \cite{hastie2005elements, geron2017hands, poldrack2019establishment}. Using out-of-sample data provides protection against overfitting and places the responsibility of model reliability in its ability to represent the natural world, as opposed to statistical significance established from in-sample data. Standard quantities may be used to evaluate model quality, such as mean squared error or mean absolute error in the case of continuous response variables; area under the receiver operator characteristic curve and the F1 score for binary responses (e.g., for logistic regression); or the Integrated Brier score and C-index for time-to-event data (survival analysis).

A good model fits the training data well, does not diverge much in fit when applied to test data, and has a low complexity. High complexity models typically have a large number of features and are algorithmically complex \cite{hastie2005elements}. They may also suffer from overfitting, with the characteristic situation of small training error and large test error.


\subsection{Explanation}

Once a final model has been determined (see Fig. \ref{fig:framework}), the model can then be used to provide explanations for the prediction. This can include traditional statistics such as odds ratios and effect sizes that are based on in-sample calculated model coefficients. Explanation can also use test data to calculate partial dependence \cite{friedman2001greedy} or determine explained variance using recursive feature elimination (e.g., \cite{young2018using, aiken2019modeling}).

\subsection{Using the framework}

Ultimately, this framework is only a guide that can be applied and adjusted for each project. For example, a longitudinal study with multiple measurements might require substantial changes in how data are managed and how a model is evaluated. However, several constants should be true for any study that includes predictive statistical modeling. The encoding of data, transformation of data, and handling of missing data should be explicitly articulated. Statistical model evaluation should use out-of-sample data to evaluate model performance. Multiple types of statistical models should be compared and evaluated before presenting results from the model. In the following sections, we demonstrate how to use this framework with two case studies.

\section{Illustrating the framework}

We have applied this framework to the study of two separate, but common studies in PER. The first study uses the LASSO data set \cite{nissen2018participation}. The LASSO data set is a multi-institution, multi-course data set that includes pre/post concept inventory responses along with other variables. Student post-scores are predicted using regression models. The second study uses the Colorado Learning about Science Survey for Experimental Physics (E-CLASS) \cite{zwickl2014epistemology} to examine differences between students in introductory physics laboratory courses and lab courses that are ``beyond the first year'' (BFY) of college.

In both cases, the data that are presented have a complex structure, which justifies trying complex models. By complex, we mean that the data may contain nonlinearities, hierarchical effects, and other peculiarities that will lead to poor functional approximation using simple methods such as ordinary least squares. For the LASSO study, first, we try the traditional ordinary least squares solution to linear regression. Second, we attempt to fit a hierarchical linear model using the restricted maximum likelihood method \cite{raudenbush2002hierarchical}. Finally, we use a gradient boosted trees model \cite{chen2016xgboost}. For the E-CLASS study, the problem is framed as a binary outcome variable. Instead of predicting the post-score like before, we predict whether the students are in an introductory course or in a BFY course. In this case, we attempt to solve the logistic regression problem in three ways: 1) with a maximum likelihood estimate method and no hierarchical effects, 2) a maximum likelihood method with hierarchical effects, and 3) a gradient boosted method without hierarchical effects. We present these two separate studies as demonstrations of the framework for statistical models that attempt to predict continuous outcome variables and binary outcome variables.

In the following section, we briefly discuss how each algorithm approaches the solution. The methods for each study, such as how the LASSO data or E-CLASS data are transformed, are then discussed in their respective sections.

\subsection{Statistical models}

In this study, we use three different methods to derive a regression model. The model is used to estimate/predict the value of a response y, denoted as $\hat{y}$, based on some covariates/predictors x. In the LASSO study, $\hat{y}$ is the posttest score. In the E-CLASS study $\hat{y}$ is a binary variable denoting whether the student is a in a BFY course or in a introductory physics lab course.
In the LASSO study, the first method we use is a traditional ordinary least squares linear regression. The second is a two-level hierarchical model \cite{raudenbush2002hierarchical}. The third is a gradient boosted trees model using the Xgboost implementation \cite{chen2016xgboost}. In the second study (E-CLASS), we instead use logistic regression models since the outcome variable is whether the student is in an introductory course or a BFY course. Below is a short description of how each method approaches the solution. Then, we describe how the models are evaluated.

\subsubsection{Ordinary Least Squares Regression}

Ordinary least squares (OLS) regression is a popular method used across all fields of science having its roots in the 1800s \cite{stigler1986history}. Ordinary least squares estimates 
a vector of coefficients for the model:

\begin{equation}
    y = \beta_{0} + \sum_{j=1}^{p} X_{j} \beta_{j} + \epsilon
\end{equation}

 by minimizing
the residual sum of squares\cite{hastie2005elements},

\begin{equation}
    RSS(\beta)=\sum_{i=1}^{N}(y_{i}-\hat{f}(x_{i}))^{2}
\end{equation}
where $y_i$ and $x_i$ are the value of $y$ and $X$ for the observation $i$, $(i = 1\dots,N)$, respectively, and $\hat{f}(x) = \hat{\beta}_{0} + \sum_{j=1}^{p} X_{j} \hat{\beta}_{j}$. This vector, $\hat{\beta} = (\hat{\beta}_1, \dots, \hat{\beta}_p)$, can be obtained directly from the unique solution (not derived here) $\hat{\beta}=(X^{T} X)^{-1} X^{T} y$, while $\hat{\beta}_0 = \frac{1}{N}\sum_{i=1}^{N}y_i$. In this case, $\epsilon$ is assumed to be the irreducible error, i.e., the variability that cannot be explained by the model.
It is also assumed that this error is normally distributed with a mean 0 and standard deviation $\sigma^2$ ($\mathcal{N}(\mu=0, \sigma^2)$).

The coefficients of an ordinary least squares model represent how a model variable can impact the outcome variable in two ways: negative/positive (the coefficient sign), the predicted shift in the outcome variable based on a one unit increase in the covariate (the coefficient magnitude).

\subsubsection{Hierarchical linear model}

Hierarchical linear regression models (also called mixed effects models) are another popular method that has seen much use in social science \cite{raudenbush2002hierarchical} and in PER (e.g., \cite{PhysRevSTPER.9.020109, PhysRevSTPER.10.020101}). Mixed effects models allow the inclusion of random effects into the classical linear model, usually something related to clusters of observations. For example,
the effect of attending a Catholic school versus a public school may have on math learning \cite{raudenbush2002hierarchical}. These regressions typically have nested equations that represent the different organizational effects described by \cite{bryk1989toward}. The simplest version of a random effects model is $y = X \beta + Z u + \epsilon$, where $X$ is the input data, $\beta$ is the fixed effects, $Z$ is input data with random effects, $u$ is a random effect ($E[u]=0$, $Var[u]=\sigma$), and $\epsilon$ is the random error. These models cannot be solved analytically like ordinary least squares and iterative methods must be used instead. For a PER treatment of these types of models see \citet{van2019modernizing, natasha2020}. 

\subsubsection{Gradient Boosting}\label{boosting}

The idea of boosting is to estimate the function $f(X)$ that relates the covariates to the response in an iterative way. At each iteration, called the boosting step, the model is improved by fitting a base estimator $h(X)$ to the negative gradient of a loss function (e.g., the squared loss). The loss function is a function that penalises the discrepancies between the outcome of a model and the truth. In the simplest case, this means to iteratively fit a model to the residuals of the model computed in the previous step. At each step, the improvement is kept small (a hyperparameter, called boosting step size, guarantees this) to allow a better exploration of the covariates and, more importantly, to avoid overfitting. Examples of boosting algorithms are the model gradient boosting \citep{hofner2014model} implemented in the R package \texttt{mboost} and the eXtreme gradient boosting \cite{chen2016xgboost}, implemented in \texttt{xgboost}. The latter, in particular, uses, as default, statistical trees \cite{hastie2005elements} as a base estimator. 

A full description of gradient boosting can be found in \cite{hastie2005elements}. The specific optimizations that Xgboost enhances gradient boosted trees with are described in \cite{chen2016xgboost}.

\subsection{Model Evaluation}

It is important to evaluate model performance in order to gauge the confidence we should place in models. Models that have poor fit, such as when $R^2$ is close to zero, may produce statistically significant results that are not repeatable or are simply due to increases in the size of modern data sets \cite{meehl1967theory}. 
Model performance for the LASSO study is evaluated visually by comparing predicted distributions (Fig. \ref{fig:lasso_pred}) and through the following metrics (Tab. \ref{tab:fitstats}): the mean square error (MSE) and the mean absolute error. Model performance for the E-CLASS study is evaluated by comparing Receiver Operator Characteristic (ROC) curves (Fig. \ref{fig:roc}). All metrics are computed on a test set.

Here, we do not explore information criteria such as the the Akaike Information Criterion (AIC) and the Bayesian Information Criterion (BIC). These criteria try to evaluate the prediction error by adding a term to the training error that is supposed to estimate the optimism, i.e. how much the training error underestimates the test error \cite{hastie2005elements}. While accurate for simple models, these criteria tend to be hard to compute for more advanced methods, such as boosting, due to the difficulties in estimating the effective degrees of freedom \cite{mayr2012importance}. $R^2$ is also not applicable here, as it often has different meaning for different models. Think, for example, to a random effect model where the random effect controls some of the variance of such an effect. Since $R^2$ represents the explained variance of the model, and the random effect model controls some of the variance through the random effect, then the $R^2$ will not capture the same explained variance behavior. Despite being a reliable measure in the linear regression model, $R^2$ should not be used for comparing models different in their nature.

Additionally, we examine the out-of-sample residuals for each model (Fig. \ref{fig:lasso_resid}). Residuals are the difference between the true value and the predicted value from a model. While most summary statistics provide average views of how good the model is in predicting the out-of-sample data (e.g., mean squared error), residuals plots allow for the examination of local behavior across the range of the outcome variable (i.e., the post-test score).

An assumption of the ordinary least squares regression model is that residuals should be normally distributed and not correlated. When the residuals are not normally distributed and/or they are correlated, this indicates the model missed some behavior in the data. That is, there is an effect that is occuring in the data causing, for example, heteroskedasticity that is not captured by the model's functional approximation of the data. While this assumption is relaxed for other types of models, the presence of correlation within residuals can be useful in analysing model outputs. Thus, visual inspection of the residuals is often important to determine if models miss underlying structure in the data or overfit on underlying structure. In practice, it is common that residuals have some correlation in real world data \cite{hastie2005elements}.

The gradient boosted models presented in this paper use decision tree estimators instead of linear models. Decision tree models do not return coefficients that are easily interpretable like in the case of linear models, instead they return the gain in information due to the use of that variable in the tree. For a gradient boosted model that is a collection of many trees, the average contribution of all the instances of the variable is used to describe its ``importance'' in the model prediction. For an xgboost model, this value is unitless \cite{chen2016xgboost}. Gradient boosted models are evaluated in the same way as linear models using out-of-sample fit statistics.

\section{Case study 1: Predicting Post-Test Scores in Concept Inventory Data}

Using the LASSO data set  \cite{nissen2018participation}, our first investigation looks at predictive regression models of post scores on the Force Concept Inventory and the Force and Motion Conceptual Evaluation. The LASSO data set is a collection of pre/post concept inventory responses collected from multiple institutions typically in introductory physics courses. The data in this paper cover three years of data collection from 2016 to 2018. The LASSO data set is connected with the national Learning Assistants' program that aims to include undergraduate students in the instruction of courses \cite{otero2010physics}. Because this program is teaching focused, LASSO data are likely to be biased by the institutions that select to participate. Institutions that are more likely to invest time and money in teaching and learning are more likely to participate in the LASSO program. Institutions that invest in teaching and learning are likely to see larger learning gains than those that do not \cite{hake1998interactive}. However, the LASSO data set is a large data set of validated concept inventory responses, thus, it is a good test bed for the methodology presented in this paper.

\subsection{Concept Inventory Data Methods}

\subsubsection{Data Management}

\begin{table}[]
\caption{\label{tab:lasso}Self-reported gender, race, and ethnicity of students in the LASSO data set. Students may select more than one race option on the survey. In this table the race categories are presented as a sum of the reported within category. Thus the total sum of reported races is greater than the total number of students (N=6688) due to students identifying as more than one race.}
\begin{tabular}{l|l}
\hline \\
 & \textbf{Number of students} \\
\hline \\
\textit{Gender} & \\
Female & 2422 \\
Male & 4266 \\ \\
\hline\\
\textit{Race/ethnicity} & \\
Hispanic & 1194 \\
White & 4381 \\
Black & 81 \\
Asian & 819 \\
American Indian & 60 \\
Hawaiian or other Pacific Islander & 60 \\
Race (other) & 748 \\

\end{tabular}

\end{table}

The 2016-2018 LASSO data set has 13916 responses for 28 institutions. We restrict these data to responses for only the Force Concept Inventory \cite{hestenes1992force} and the Force and Motion Conceptual Evaluation \cite{thornton1998assessing}. This focuses our work on introductory mechanics concepts.  We then remove all courses with less than 20 students enrolled and all courses with less than 20 responses. This increases the registered number of students per course for the training and testing data sets. We also remove data for any course that has less than 50\% students responding to the pre-test or the post-test. Additionally, for students with multiple responses, we use only the student's first response. This reduces the data set to have 6688 students total across 20 institutions and 90 courses.

In this paper, we have restricted the total variables to 30, including the outcome variable, the posttest score. The variables include two data types: (1) data regarding the student, and (2) data regarding the course the student attended (see Tab. \ref{tab:datatypes}). Student data include the pretest score, background information for students, such as gender and race/ethnicity, the number of pre and post test questions answered, what year the student is in while taking the course (1st, 2nd, or 3rd or greater), and the time the student spends taking the pre and post test. The course variables include the type of concept inventory (FMCE or FCI), the number of students enrolled in the course, the number of learning assistants assigned to the course, the student-to-learning assistant ratio, and whether the course is offered at a community college or not.

The LASSO data set has missing data \cite{nissen2019missing}. Pre-test scores, post-test scores, the number of questions answered for pre and post-tests, and the time it takes to complete a pre and post test all may have missing data. To address this issue, we used the Multivariate Imputation by Chained Equations (MICE) method \cite{buuren2010mice} as recommended by \cite{nissen2018comparison} for use on the LASSO data. MICE estimates the missing data per variable that has missing data via a regression algorithm using all other columns. This is repeated for each column with missing data until all columns are filled. This is then iteratively fit across a user chosen maximum number of iterations. In this paper, MICE is used to impute the missing data for pre-test scores, post-test scores, the number of questions answered for pre and post-tests, and the time it takes to complete a pre and post test. For each variable that includes imputed data, an additional column is created to identify whether the data are imputed or not. This additional column can, in the case when data are not missing completely at random, provide models with information that can aide prediction \cite{josse2019consistency}.

\subsubsection{Model Evaluation}

In the LASSO study, we use an OLS model, a mixed effects model, and a boosted model. The OLS model uses all of the variables found in Tab. \ref{tab:datatypes}. The mixed effects model additionally considers a random effect on the intercept and on the slope. The boosted model uses the xgboost package in python \cite{chen2016xgboost} and uses decision trees as estimators. It uses 20000 decision tree estimators. 20000 was chosen as it provides a large number of estimators to learn the data without having any one estimator overwhelming the fit. We choose these three models as they each attempt to solve a specific problem and the OLS and mixed effects models are becoming commonly used in PER \cite{van2019modernizing}. OLS models are used to estimate mean effects on outcome variables (in this case, the concept inventory post score). Mixed effects models are used when there is an assumed group-wise variance such that the slopes and/or intercepts per group may be different \cite{bryk1989toward, van2019modernizing}. Gradient boosting models have been developed more recently  \cite{friedman2002stochastic} and can often better fit data in comparison to maximum likelihood models, such as the mixed effects models presented in this paper  \cite{boostingl2loss}.

\begin{figure}
    \centering
    \includegraphics[scale=0.75]{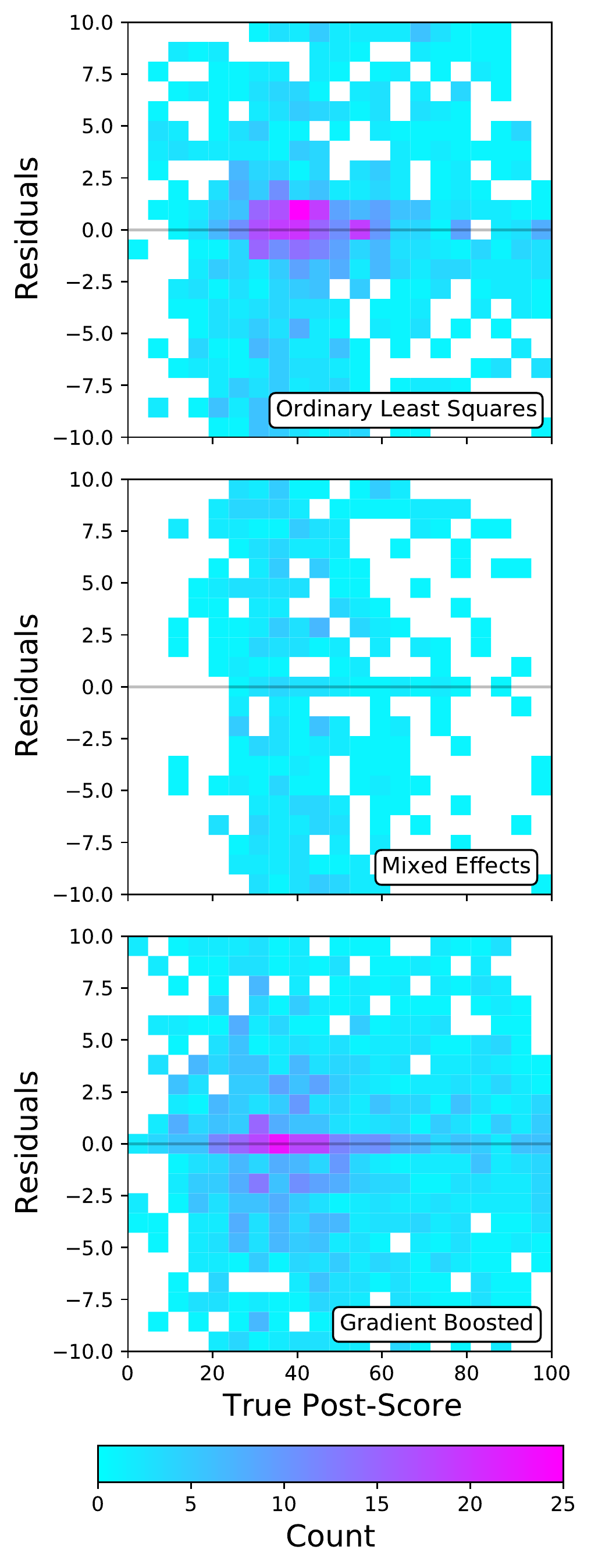}
    \caption{Residuals plots for the LASSO study. For the OLS model and the xgboost model, the residuals are normally distributed with the majority being within $\pm 10$ points of the true value (see appendix Fig. \ref{fig:residlassodist}). The mixed effects model does not produce normally distributed residuals.}
    \label{fig:lasso_resid}
\end{figure}

The models are then evaluated using the visual predicted post-score distribution (Fig. \ref{fig:lasso_pred}), the residuals visual (Fig. \ref{fig:lasso_resid}), the mean squared error, and the mean absolute error (Tab. \ref{tab:fitstats}). A visual examination of the model fit through predicted post-score distributions and residuals provides a visual representation of the fit statistics (such as mean square error). This visual examination can prevent us from being ``fooled'' by good fit statistics \cite{anscombe1973graphs}. The fit statistics provide a summary value that can be compared directly across models. In this case, the lower the mean squared error and the mean absolute error, the better fit the model.

\begin{table*}[]
\caption{Data Types at different levels (student or course level). An asterisk (*) indicates that there is also a column in the data set indicating if the data was imputed or not. No columns for the E-CLASS study were imputed.}
\begin{tabular}{l|lll}
\hline \\
\textbf{Data Type} & \multicolumn{1}{l}{\textbf{LASSO Study}} & \multicolumn{1}{l}{\textbf{E-CLASS Study}} \\
\hline \\
\textbf{Student Level} & \begin{tabular}[c]{@{}l@{}}Pre-test Score*\\ Post-test Score*\\ N\% Post Answered*\\ N\% Pre Answered*\\ Log Pre-Duration*\\ Log Post-Duration*\\ Race\\ Hispanic\\ Gender\\ Year Enrolled\end{tabular} & \begin{tabular}[c]{@{}l@{}}Race/Ethnicity \\ Gender\\ Responses to career goals Q's\end{tabular}\\
\hline \\
\textbf{Course Level} & \begin{tabular}[c]{@{}l@{}}FCI or FMCE\\ Community College\\ Pre-tests Completed\\ Number of LAs\\ Post-tests Completed\\ Student to LA Ratio\\ Students Enrolled\\ Course ID\\
\end{tabular} & \begin{tabular}[c]{@{}l@{}}Institution type\end{tabular}
\end{tabular}
\label{tab:datatypes}

\end{table*}

\subsection{Concept Inventory Data Results Communication}

\begin{table*}[]
\caption{Out-of-sample model fit statistics for ordinary least squares, mixed effects, and gradient boosted models for the LASSO data study. In all cases, the gradient boosted models have  ``better'' fit statistics in comparison to the linear models. This better fit is unlikely due to the boosted model over fitting the data. This better fit is evidenced by the linear models producing significantly different post score distributions using the out-of-sample data (see Fig. \ref{fig:lasso_pred}).} 
\begin{tabular}{lccc}
\hline

 \textit{LASSO Data} & \textbf{Ordinary Least Squares} & \textbf{Mixed Effects} & \textbf{Gradient Boosted} \\
\hline
 \textbf{Mean Absolute Error} & 0.0975 & 0.111 & 0.0804 \\
 \textbf{Mean Squared Error} & 0.0200 & 0.0217 & 0.0152\\

 \hline \\
\end{tabular}

\label{tab:fitstats}
\end{table*}

Using the LASSO data set, we have trained three separate models: 1) an ordinary least squares model, 2) a mixed effects model, and 3) a gradient boosted model. Overall, the gradient boosted model has a 
lower mean absolute error and lower mean squared error than the linear models (Tab. \ref{tab:fitstats}). The mixed effects model has slightly better fit statistics in comparison to the ordinary least squares model.

\begin{figure}
    \centering
    \includegraphics[scale=0.34]{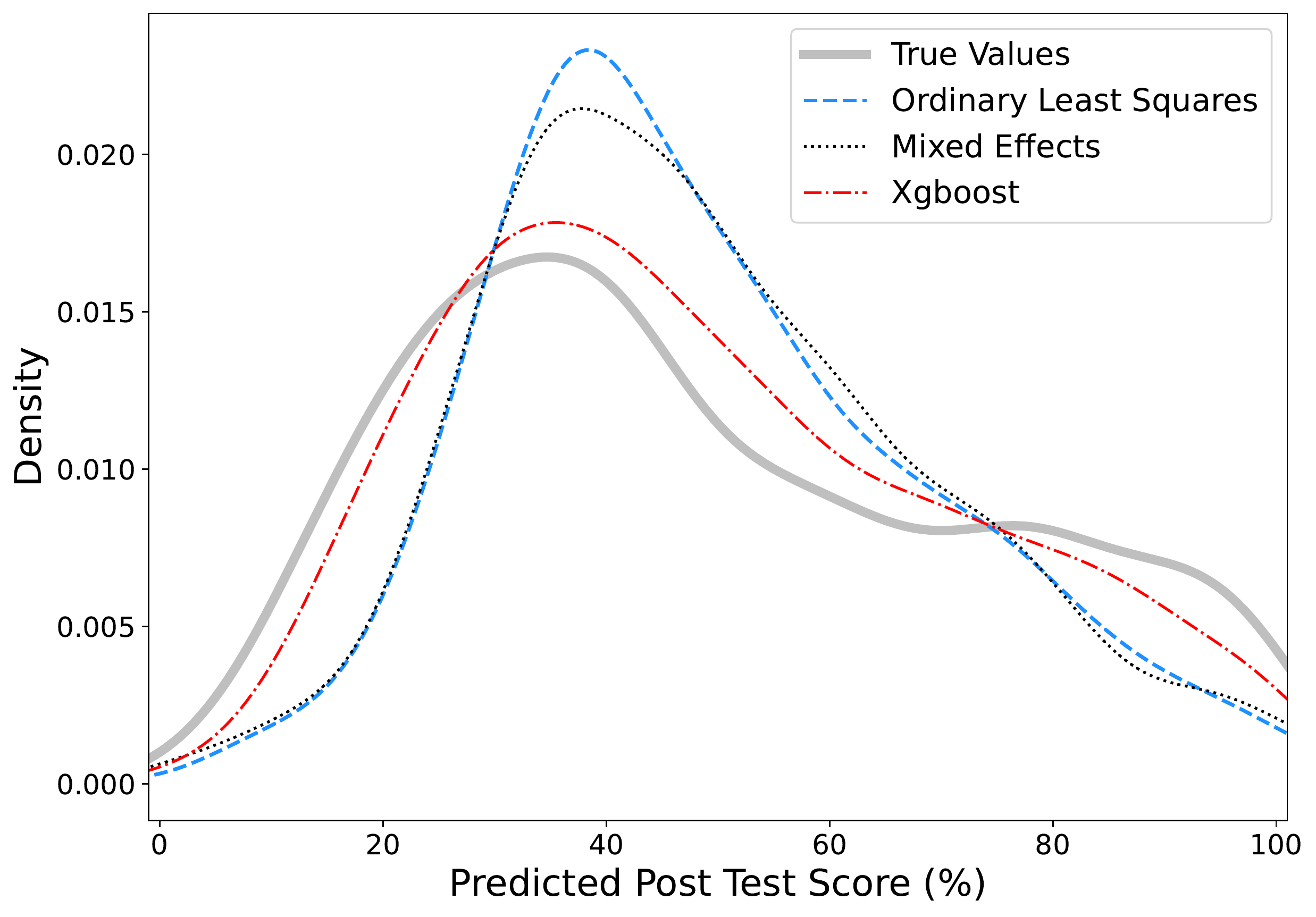}
    \caption{A comparison of the out-of-sample predicted distributions of post test scores from the LASSO data set. Curves represent the kernel density estimates of the predicted post-test scores. The gradient boosted model is closer to the true distribution than the ordinary least squares model and the mixed effects model. 
    }
    \label{fig:lasso_pred}
\end{figure}

Visually, the out-of-sample post-score distribution predicted by the gradient boosted model and the mixed effects model is close in shape to the true values (Fig. \ref{fig:lasso_pred}). 
The ordinary least squares and hierarchical models over predict scores in the 30-60\% range and under predict higher scores. 

The residuals for all models show some correlation with the true values of the post-test score. The ordinary least squares model under predicts low scores by a wide margin in comparison to the mixed effects model and the gradient boosted model. The ordinary least squares model and the gradient boosted have larger numbers of very small residuals in the range of the most likely scores (20\% to 60\%). The mixed effects model has a much larger spread of residuals in this region.

\begin{figure*}[!ht]
    \centering
    \includegraphics[scale=0.7]{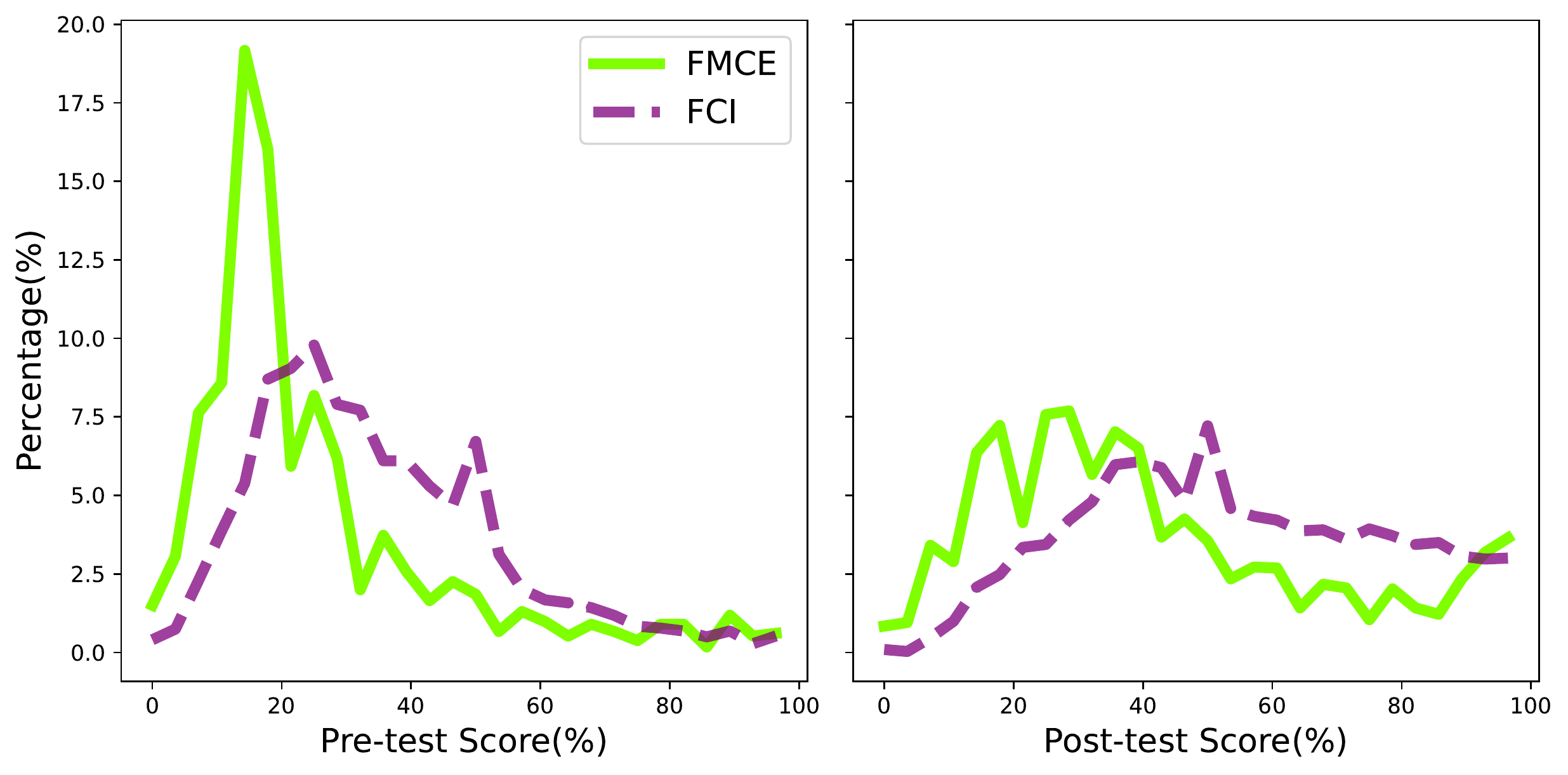}
    \caption{Distribution of pre and post scores for FCI and FMCE courses from LASSO data set (N=6688). In both the pre and post score cases, the distribution of scores are right skewed normally distributed. Data come from multiple institutions across multiple courses ($N_{courses}=90$).}
    \label{fig:lasso_dist}
\end{figure*}

\section{Case Study 2: Comparing students in introductory vs beyond first year courses}
Not all models in PER attempt to predict a continuous dependent variable like we have shown in the previous analysis. In some cases, the dependent variable is binary or even multi-class. For this work, it is common to use logistic regression instead of the traditional regressions. In this study, we use a subset of the data E-CLASS to present an example of a binary classification task in PER and how the framework can be used for such a task. This subset focuses on the future plans questions asked of students in the E-CLASS post survey.

\subsection{E-CLASS methods}

\subsubsection{Data Management}

\begin{table}[]
\caption{Self-reported gender, race, and ethnicity of students in the E-CLASS data set separated by level of course (First Year and Beyond First Year).}
\begin{tabular}{p{5cm}|ll}

\hline \\
& \textbf{Intro.} & \textbf{BFY} \\
\hline \\
\textit{Gender} &  &  \\
Women & 3613 & 463 \\
Men & 5138 & 829 \\
Other & 91 & 14 \\ \\
\hline\\

\textit{Race/ethnicity} &  &  \\
American Indian or Alaska Native & 13 & 2 \\
Asian & 1738 &	217 \\
Black &	380 &	31 \\
Hispanic/Latino	& 841 &	48 \\
Multi-race & 863 & 100 \\
Native Hawaiian or other Pacific Islander &	31 &	1  \\
Not Reported &	542 &	107 \\
White & 4434 & 800 \\
\end{tabular}

\end{table}

The E-CLASS is a survey of student epistemology and expectations surrounding experimental physics. We take a situated theory of epistemological development~\cite{Sandoval2014} approach when considering how the survey is used and what research questions can be asked. This theory looks at how  students' epistemologies develop in particular social-cultural contexts based on reflections of their experiences.  The survey is validated for all levels of university/college students and can be applied across many different lab experiences. To facilitate this process, the E-CLASS survey is administered through a central administration system \cite{wilcox2016students}. As of January 2020, it had been used in 599 classes across both introductory lab courses and BFY courses.
Data was reduced or removed for several reasons. Students were able to reply to the gender question as a text response instead of picking an option. In this case, these responses were reduced to a third ``other'' category. In some cases students had replied to the survey more than once, e.g., they took the survey in more than one course. In this case we only used their first response. In some cases there was no description of the type of institution the student attended in the course information survey that instructors filled out. Responses were removed in this instance. Additionally, responses were removed for students who attended 2 year colleges since there are no BFY courses in these institutions. This reduced the overall data set of matched pre/post responses of N=19445 to N=10148 students. In this study, we focus on the differences between students in introductory and BFY courses.

In this study, we do not use any results from the attitudinal survey itself. Instead, we use demographic information and some of the responses to career plan questions that the students report at the end of the survey. This includes the student's gender, eight questions (see appendix) concerning the student's preferred career trajectory, the student's race/ethnicity, and the type of institution the student attends (4 year college, Masters granting, or PhD granting). We then use these variables to predict whether the students are in introductory or BFY courses.

\subsubsection{Model Evaluation}

We apply traditional maximum-likelihood logistic regression, a mixed-effects logistic regression, and a gradient-boosted logistic regression to compare students in introductory courses to those in BFY courses. We compare the predicted output using the receiver operating curve (ROC) \cite{hosmer2013applied}. This curve represents the ratio of true-positive to false-positive predictions of the model given the entire scale of the decision boundary (0 to 1). For example, a threshold of 0.2 would indicate that any student with a predicted probability of $<0.2$ would be classified as a student in an introductory course. Students above this line would be classified as BFY students. For the hierarchical model, we add a random term to handle the effect of the institution type.

In this study, we used the advanced form of the ``train-test-split'' method presented in Sec. \ref{sec:traintestsplit} known as K-fold cross validation \cite{hastie2005elements}. Using 5 folds, we calculate each model and then report the average Area Under the ROC (AUC) curve(Fig. \ref{fig:roc}). By using cross-validation, we can develop a better understanding of our confidence in the comparisons of the reported AUCs. This is necessary because a) they end up being low according to acceptable AUC ranges created by \citet{hosmer2013applied}, and b) they are close in value to each other.

\subsection{E-CLASS Results Communication}

\begin{figure}
    \centering
    \includegraphics[scale=0.42]{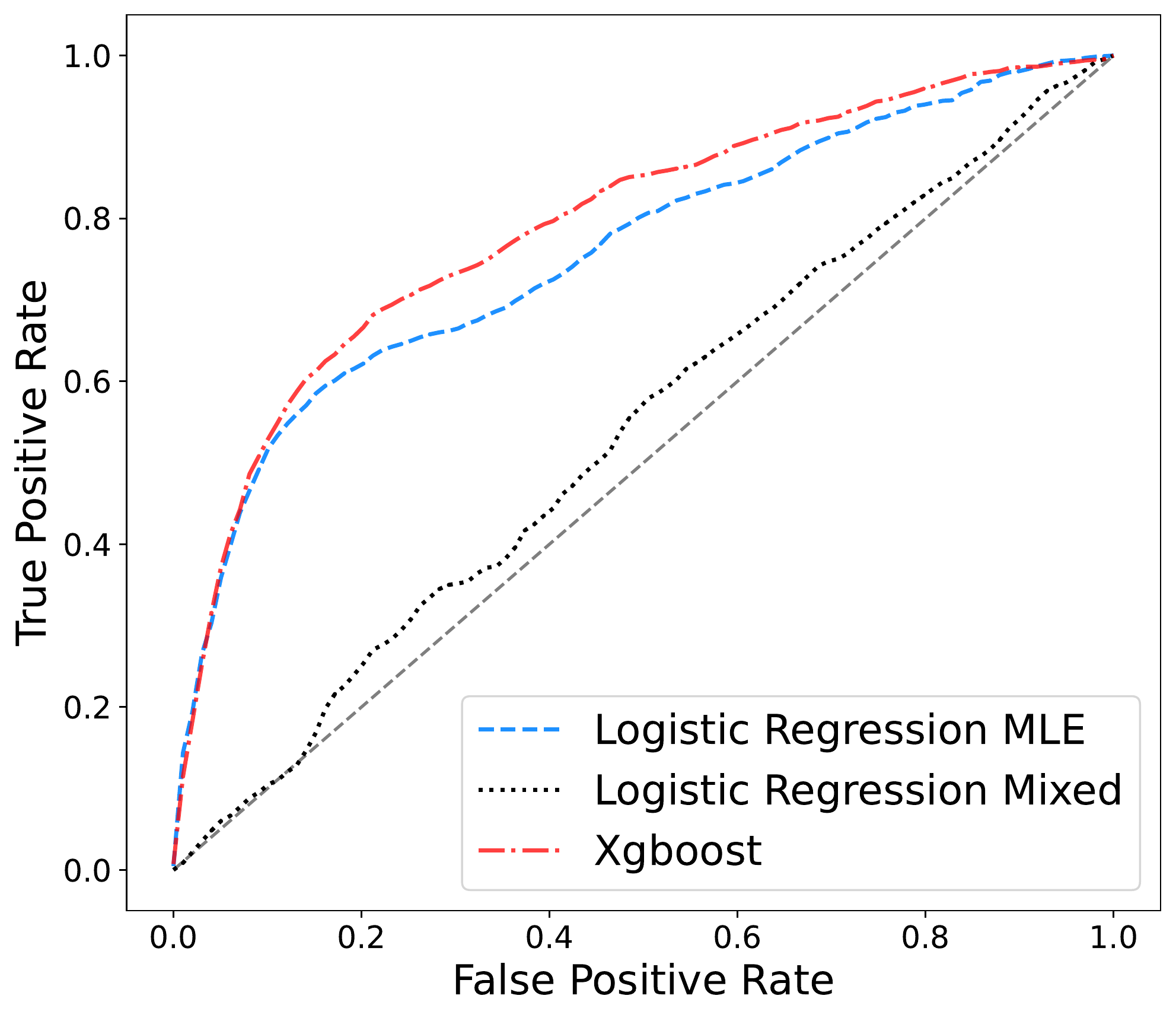}
    \caption{Receiver Operator Characteristic (ROC) curves for the E-CLASS models predicting whether a student is in an introductory course or a BFY course. The more area under a curve, the better its performance. Overall, the boosted model (AUC=0.798) performs approximately 2\% better than the maximum likelihood model (AUC=0.773) and considerably better than the mixed effects model (AUC=0.581). However, the overall scores are still below an ``excellent discrimination'' (AUC$>$0.8) as identified by \citet{hosmer2013applied}. In this figure, the curves represent the average ROC curve for all 5 cross validation folds.}
    \label{fig:roc}
\end{figure}

Using the E-CLASS data set, we trained three models to predict whether a student would be in an introductory course or a BFY course when responding to the survey. Using only demographics, responses to the seven-question career-plan question block, and information about the type of institution, we calculated the ROC curves shown in Fig. \ref{fig:roc} from the model classifications (True Positive Rate and False Positive Rate). Overall the gradient boosted logistic regression model performs the best (AUC=0.798),  while the mixed effects logistic regression model performed the worst (AUC=0.581). The fixed effects logistic regression model performs similar (AUC=0.773), but slightly worse, as compared to the gradient boosted model.

All three models are below an ``excellent discrimination'' (AUC$>$0.8) as identified by \citet{hosmer2013applied}. It is not clear if this can be expected. One must consider the typical demographics of physics lab courses in the US. Introductory lab courses often have students in wide-ranging majors, from physics, to engineering, to those in the life sciences, while BFY courses are typically dominated by physics majors. Students in physical sciences and engineering in the US tend to be overwhelmingly white and male, while those in the life science majors have more women. Additionally, there are issues of systemic sexism and racism in colleges and universities that can decrease the percentage of women and non-white students continuing into upper-division courses \cite{systemicscherr2017}. It is not clear \emph{a priori} if these general differences in course population are seen the in the data we have or can be identified by the models.

\section{Discussion}

The framework that is presented in this paper is a culmination of a number of contributions towards statistical modeling in PER. It highlights the need to understand the types of data we collect, how we can represent them, and how we can transform the data \cite{springuel2019}. The framework is attentive to the missing data, which is common in social science \cite{nissen2019missing}. The framework allows for the introduction and evaluation of many different kinds of statistical models that can elucidate understanding of educational data \cite{van2019modernizing, aiken2019modeling, zabriskie2019using}. Finally, the framework aligns statistical model evaluation in PER with calls in social science fields to place a much greater emphasis on evaluating model prediction \cite{poldrack2019establishment}. 

\subsection{Data Management}

Describing data sets used in PER includes how the data are managed. This includes explicit description of how the data are reduced (e.g., dropping data that could be used to identify students) or describing how data are transformed (e.g., how the durations in the LASSO study were transformed via a logarithm). Describing this process allows for interpretation and reproducibility of transforming the raw data into data for a statistical model.

It is common in the social sciences that data are missing and that we need to impute that data in some way \cite{simons2015multiple, jakobsen2017and, hayes2017should, nissen2019missing}. The reason data are missing can be due to both completely random reasons, such as a question simply not being answered on a survey, to purposeful reasons, such as a sub-population under study is less likely to participate in a repeated measurement like a pre/post test.

In the LASSO data study, we imputed missing data. Examining the feature importances, The gradient boosted model used many of the ``is imputed'' variables more effectively for prediction than it did the actual data (see Appendix). Approximately half (47\%) of the rows in the LASSO data set had at least one column that was imputed. It could be that having so much data being imputed biases the model much like not imputing data at all can bias results \cite{nissen2019missing}. That is, when so much data are being imputed, the model learns the process the imputation engine follows and that can artificially inflate the model fittedness. Thus, the statistical model then represents what the imputation model did, instead of nature of the data collected. It may be that there are too many rows with missing data. This is similar to recommendations from biomedical clinical research, which state not to impute data when data are missing at least 40\% of the time \cite{jakobsen2017and}. It could also be that while the overall LASSO data set is large, group-wise per course and institution, it has low N values. That is, in both the student groups (e.g., gender, race) and the institutional groups (e.g., single courses) there is a low number of students that represent each sub-population. Thus, the actual amount of data available to impute student data from the same learning community is low. In this case, it may be wise to examine other methods such as weighting variables instead of imputing variables that have shown to be effective in low-N scenarios \cite{hayes2017should}.

\subsection{Model Evaluation}
\label{sec:discmodeval}
We have presented two research studies in which we demonstrated the utility of examining model fittedness using the statistical model framework presented in this paper. This framework takes an agnostic approach to understanding model success. It is agnostic to whether models are considered direct representations of the data or are simply algorithms that fit data. There is much contention in statistics as to which viewpoint is ``better.'' \citet{bryk1989toward} argued for directly representing the organizational levels of an educational environment through random effects in statistical models. This argument forms the basis for the use of most hierarchical mixed effects models. These models have been demonstrated to reject previous results such as the interaction of aptitude and teaching style has on math performance \cite{cronbach1975between}. Ultimately, we must pick the best models that can explain relationships we are interested in, being careful of not searching for the model that explain it by chance. Picking these models must rely on a comparative assessment of many models performance \cite{poldrack2019establishment}. In many cases, there may be a ``multiplicity of good models'' \cite{breiman2001statistical}. 

In the LASSO study, we assume there is an institutional random effect due to them being reported in the literature (e.g., \cite{nissen2018participation}). Therefore, the ordinary least squares method performs poorly and there is an assumed random effect due to the institution. That is, there is not only a per institution effect that adds some amount to a student's total post-test score, but also this effect has a wide variance that would make the effect variable per student all other variables being equal. 

For the E-CLASS study, we used the Receiver Operator Characteristic Curve (Fig. \ref{fig:roc}). The ROC curve represents the ratio of true-positive results to false-positive results. Thus, it helps us estimate model performance in the context of what the model labels as the true value (in our case, being a BFY student). What we see in this study is that if we had relied on an assumed structure of the data as dictated by the mixed effects model, we would arrive at much different results than either the OLS model or the boosted model.

\subsubsection{Visualizing model results}

We have also emphasized in this paper the visual examination of model fits to data, in addition to the fit statistics. It is not a new claim that we must not only examine model fit statistics, but also visually examine the model's predictions themselves. \citet{anscombe1973graphs}'s quartet highlighted the need to examine the data visually. Using a quadrant of four different data sets but with identical fit statistics, Anscombe demonstrated that visual examination of a model's fit is sometimes the only way to determine our trust in the model itself. In this paper, we have demonstrated the same need when statistical modeling is applied within physics education research. In the LASSO case, the gradient-boosted model arrived at a better understanding than the linear models of the data set. The residuals of the model were examined to better understand the model fit. Using the residuals we can determine how closely the predicted scores are to the actual scores.

\subsubsection{Boosting}

We have presented three models for each study: (1) a traditional linear model, (2) a mixed effects model, and (3) a boosted model. In each case, the boosted model is better at fitting the data without overfitting. The boosted method likely solves the regression better because the model spends more time learning the peculiarities of each data set \cite{boostingl2loss}. Gradient boosted models are typically fit using a large number of iterations of adding new learners to the overall base model. The number of iterations is actually an important tuning parameter for boosting, and many properties depend on its correct choice \citep{mayr2012importance}.

The models presented here use 20000 iterations, which produces a post-test distribution closer to the observed (test) data and an overall better fit. No one learner in the boosted model uses all of the variables nor does it use all of the data. By combining the collection of many weak learners, the boosted model is able to capture complex structures of the data, i.e., non linear effects, and prevents overfitting (the typical sign of overfitting a model is when the model predicts training data very well and performs poorly on test data \cite{hastie2005elements}). In particular, the last point is related to the discussion of Section \ref{sec:discmodeval} on the role of a model: if one wants the the results to be generalized outside the specific data set at hand, i.e., have a low prediction error, a boosted model may lead to a much better model than those obtained with traditional approaches.

Ultimately, it is up to the researcher to decide what models they use and why. Researchers who are most concerned with understanding local effects and not producing generalized models are likely to find ordinary least squares an adequate solution for their purposes. However, by restricting the use of which models are available, it might lead to models that require lower confidence in the results or even erroneous characterization of the most important features. Models may also not be able to be generalized beyond a local context. Newer methods, such as boosting, also have the ability to increase understanding of sub-populations that traditionally are poorly fit by statistical models due to lack of data \cite{aikenplosone}. Ultimately, there is not one ``true'' model, but many different models that can inform us on the dynamics of students in educational systems in physics.

\subsection{Results Communication}

It is common in PER to report statistical model results as the coefficients and random effects (if any) of the model. We argue that we can present results in greater detail by also using visualizations of the predicted output.

Models with mis-estimated variable rankings may lead to departmental policy decisions that are not reflective of the confidence researchers have in the model. Given that 24\% of the reviewed papers (Sec. \ref{sec:background}) presenting statistical models had no examination of model fittedness, this could have a profound effect on results that have been presented in physics education research. 

It is often the goal in PER that we are interested in as much the magnitude of an effect as is the direction of that effect. For example, it is helpful to know not only how much a student's pre-test score affects their post-test score, but also whether that effect is negative or positive. In this paper, we have not presented these types of results. This is done in order to focus on the examination of prediction to establish model confidence. Modern methods such as boosting can produce these estimates as well depending on the type of learner the boosted model uses (such as a linear learner). These boosted models can also use decision trees (this is what is used in the E-CLASS study). Feature importances of tree-based models do not tell us positive or negative effects only the magnitude of the effect. Instead, tree based models can utilize methods such as partial dependence \cite{friedman2002stochastic} or shapely additive values \cite{lundberg2017unified} that can provide a more complete picture of each variables effect on the outcome. Both of these model explanation methods explore the entire range of a particular variable and its effect on an outcome variable such as a post-test score. These measurements are able to explain non-linear effects (impossible for model coefficients) and are model agnostic. Explaining these methods is beyond the scope of this paper.

\subsection{The role of theory and domain knowledge in statistical modeling}

This paper has focused primarily on developing a framework of statistical modeling in PER. The studies presented as examples here are classic studies in PER: survey and concept inventory research. While there is theory motivating these types of studies, the theories were not present in this paper. This is not a tacit recommendation to avoid theory. Theory dictates the research design and the boundary conditions upon which we accept or reject model results. This holds true as well in the reverse. In some cases, theory may fail and the statistics we have used provide an argument for that failure. In either case, it is important to clearly articulate the theory that motivates a study, as much as it is important to articulate a statistical model evaluation process.

\subsection{Recommendations for Researchers}

Researchers who are building statistical models to answer physics education research questions can build more effective research by following a few recommendations:

\begin{enumerate}
    \item Describe the data set, how it was collected, the types of data that are utilized in the study, the context of the data whether it's from a specific institution or many, and any processes used to impute the data or otherwise alter it \cite{springuel2019, knaubaikending2019}.
    \item Predictive modeling studies should use out-of-sample data to evaluate model predictive power prior to relying on model output for explanation \cite{hastie2005elements}.
    \item Statistics such as the mean squared error and the mean absolute error are appropriate test statistics for comparing model output of prediction models \cite{poldrack2019establishment}. 
    
    Examining receiver operator curves (such as shown in the E-CLASS analysis in this paper) are appropriate for classification models \cite{poldrack2019establishment}.
    \item Studies should visually examine the predicted distributions and residuals of regression models. 
    \item Researchers should compare multiple models to produce the best fitted models \cite{breiman2001statistical}. These models should include not only different groups of variables, but also different algorithms such as gradient boosting \cite{friedman2002stochastic} or random forests \cite{breiman2001random}.
\end{enumerate}

While inline with current calls in other fields to standardize model evaluation (e.g., \cite{poldrack2019establishment}), these recommendations are not exhaustive. In many cases, there is likely a better fit statistic, visualization, or methodological approach than presented in this paper for the specific research question a physics education researcher is exploring. The recommendations presented here are suggested to be starting points for statistical modeling in PER.

\section{Conclusion}

We presented a framework for statistical modeling using predictive models that are presented in physics education research. The framework focuses on explicit articulation of the data management procedures, the model evaluation methods, and the way results are defined and communicated. It presents results that suggest explicit articulation of data handling through transformations, imputations, etc. is important to understand the statistical model outcomes. We highlight the need to assess statistical models using out-of-sample data. Out-of-sample prediction must be used as an assessment of the effectiveness of predictive models. Assessing prediction provides confidence in statistical model results that are otherwise lacking in view of modern statistics \cite{poldrack2019establishment}.
The framework is grounded in recent efforts to critically evaluate statistical methods in PER highlighting this need to assess prediction \cite{FocusedCollection}. We present two complementary studies that justify this need for demonstrating that without assessing prediction, researchers can arrive at erroneous results. In our two example case studies, the boosted models are consistently better than the linear models. While this is consistent with theoretical assessment of boosting from statistics literature (e.g., \cite{boostingl2loss}), the goal of this paper is not to present boosting to the PER community as a panacea. Instead it is to demonstrate the effectiveness of the presented statistical modeling framework.

\section{Acknowledgements}

This project was supported by the Michigan State University College of Natural Sciences including the STEM Gateway Fellowship and the Lappan-Phillips Foundation, the Association of American Universities, and the Norwegian Agency for International Cooperation and Quality Enhancement in Higher Education (DIKU), which supports the Center for Computing in Science Education. This project has also received support from the INTPART project of the Research Council of Norway (Grant No. 288125) and the National Science Foundation (Grant No.  PHY-1734006). We would also like to thank Coline Bouchayer who created Figure \ref{fig:framework}.


\appendix
\section{E-CLASS questions}

The future plans questions from the E-CLASS post survey used in this study include:

\begin{enumerate}
    \item Do your future plans include physics graduate school?
    \item Do your future plans include a non-academic science/math/eng job?
    \item Do your future plans include medical school?
    \item Do your future plans include other professional school?
    \item Do your future plans include teaching K-12 science/math?
    \item Do your future plans include teaching college science/math/engineering?
    \item Do your future plans include a non-science/math/engineering job?	
\end{enumerate}

\section{Model parameters}

\subsection{LASSO study}

The LASSO study presented three separate models: 1) an ordinary least squares (OLS) regression, 2) a mixed effects regression, and 3) a gradient boosted regression. The feature coefficients and importances are presented below. Additionally a plot of the residual distributions for each model is presented.

\begin{figure}
    \centering
    \includegraphics[scale=0.4]{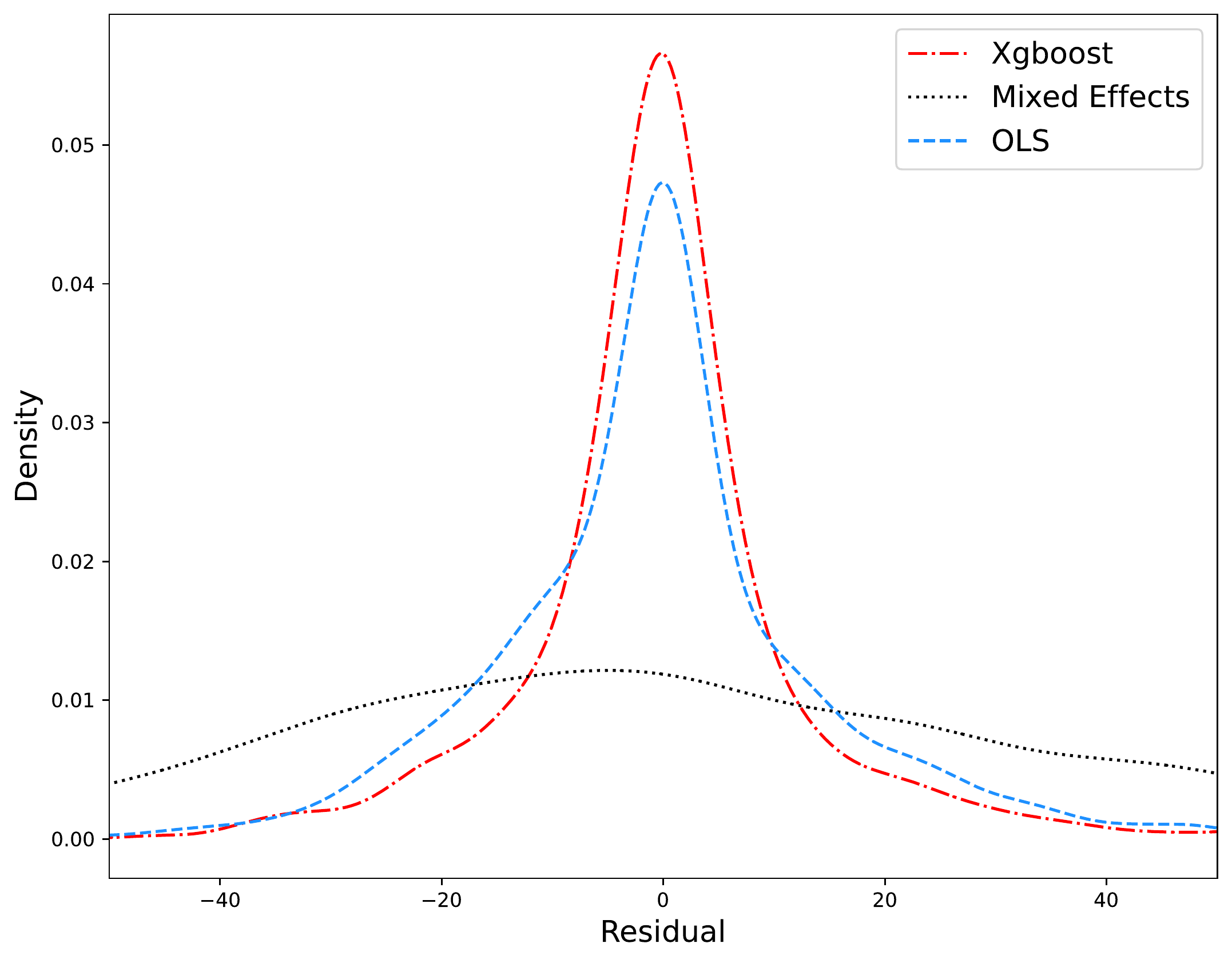}
    \caption{Residual distributions for LASSO study.}
    \label{fig:residlassodist}
\end{figure}

\begin{table*}[]
\begin{tabular}{lllllll}
\hline \\
\textbf{Variable}                & \textbf{Coefficient} & \textbf{Standard error} & \textbf{t} & \textbf{$P>|t|$} & \textbf{{[}0.025} & \textbf{0.975{]}} \\
\hline \\
Intercept                        & -33.8622      & 5.009            & -6.761     & 0.000                       & -43.681           & -24.043           \\
course\_id                       & 0.0013        & 0.001            & 0.959      & 0.338                       & -0.001            & 0.004             \\
pre\_tests\_completed            & -0.0731       & 0.011            & -6.827     & 0.000                       & -0.094            & -0.052            \\
post\_tests\_completed           & -0.1110       & 0.018            & -6.309     & 0.000                       & -0.145            & -0.076            \\
FMCE                             & -18.6029      & 2.545            & -7.309     & 0.000                       & -23.592           & -13.613           \\
FCI                              & -15.2593      & 2.508            & -6.085     & 0.000                       & -20.176           & -10.343           \\
students\_enrolled               & 0.1507        & 0.014            & 11.100     & 0.000                       & 0.124             & 0.177             \\
number\_of\_las                  & 0.1353        & 0.164            & 0.825      & 0.409                       & -0.186            & 0.457             \\
student\_to\_la\_ratio           & 36.7851       & 7.736            & 4.755      & 0.000                       & 21.619            & 51.952            \\
community\_college               & 0.4356        & 1.509            & 0.289      & 0.773                       & -2.523            & 3.394             \\
log\_pre\_duration               & -1.6383       & 0.797            & -2.055     & 0.040                       & -3.201            & -0.076            \\
log\_post\_duration              & 25.8368       & 0.728            & 35.481     & 0.000                       & 24.409            & 27.264            \\
pre\_answered                    & -0.1613       & 0.024            & -6.837     & 0.000                       & -0.208            & -0.115            \\
post\_answered                   & 0.1560        & 0.030            & 5.182      & 0.000                       & 0.097             & 0.215             \\
pre\_score                       & 0.7692        & 0.012            & 62.290     & 0.000                       & 0.745             & 0.793             \\
gender                           & -3.4316       & 0.445            & -7.707     & 0.000                       & -4.305            & -2.559            \\
hispanic                         & -4.0267       & 0.629            & -6.405     & 0.000                       & -5.259            & -2.794            \\
white                            & 5.1159        & 0.775            & 6.602      & 0.000                       & 3.597             & 6.635             \\
black                            & -1.1678       & 1.985            & -0.588     & 0.556                       & -5.060            & 2.725             \\
asian                            & 3.7404        & 0.933            & 4.009      & 0.000                       & 1.911             & 5.569             \\
race\_other\_reduced             & 0.6977        & 0.951            & 0.734      & 0.463                       & -1.166            & 2.562             \\
year\_1st                        & -2.0976       & 6.419            & -0.327     & 0.744                       & -14.682           & 10.487            \\
year\_2nd                        & -2.9017       & 6.415            & -0.452     & 0.651                       & -15.479           & 9.675             \\
year\_3rd\_or\_greater           & -1.8904       & 6.408            & -0.295     & 0.768                       & -14.453           & 10.672            \\
is\_imputed\_log\_post\_duration & -0.0805       & 0.782            & -0.103     & 0.918                       & -1.613            & 1.452             \\
is\_imputed\_log\_pre\_duration  & 1.0563        & 0.824            & 1.282      & 0.200                       & -0.559            & 2.671             \\
is\_imputed\_post\_answered      & -0.9840       & 0.440            & -2.238     & 0.025                       & -1.846            & -0.122            \\
is\_imputed\_post\_score         & -0.5320       & 0.350            & -1.518     & 0.129                       & -1.219            & 0.155             \\
is\_imputed\_pre\_score          & -0.9840       & 0.440            & -2.238     & 0.025                       & -1.846            & -0.122            \\
is\_imputed\_pre\_answered       & -0.5320       & 0.350            & -1.518     & 0.129                       & -1.219            & 0.155            
\end{tabular}
\caption{OLS coefficients for the OLS Model presented in the LASSO study.}
\end{table*}

\begin{table*}[]
\begin{tabular}{llll}
\hline \\
\textbf{Variable} & \textbf{Coefficient} & \textbf{Standard Error} & \textbf{t-value} \\
\hline \\
(Intercept) & -39.1207 & 13.47173 & -2.904 \\
log\_pre\_duration & -2.0198 & 0.76716 & -2.633 \\
log\_post\_duration & 23.84125 & 0.71204 & 33.483 \\
pre\_answered & -0.1547 & 0.02311 & -6.695 \\
post\_answered & 0.18577 & 0.02882 & 6.446 \\
pre\_score & 0.7414 & 0.01686 & 43.983 \\
gender & -2.96961 & 0.44499 & -6.673 \\
hispanic & -3.95996 & 0.61904 & -6.397 \\
white & 4.6751 & 0.75039 & 6.23 \\
black & -2.32269 & 1.95085 & -1.191 \\
asian & 2.61889 & 0.9057 & 2.892 \\
race\_other\_reduced & 0.16939 & 0.91586 & 0.185 \\
year\_1st & -4.29582 & 6.24306 & -0.688 \\
year\_2nd & -4.86221 & 6.23259 & -0.78 \\
year\_3rd\_or\_greater & -4.5968 & 6.22859 & -0.738 \\
is\_imputed\_log\_post\_duration & -1.58494 & 0.97335 & -1.628 \\
is\_imputed\_log\_pre\_duration & 0.59404 & 0.8315 & 0.714 \\
is\_imputed\_post\_answered & -0.62993 & 0.93741 & -0.672 \\
is\_imputed\_post\_score & -0.92841 & 0.68935 & -1.347 \\ \\
\textbf{Group} & \textbf{Variable} & \textbf{Variance} & \textbf{Standard Deviation} \\
\hline \\
course\_id & (Intercept) & 1.48E+01 & 3.84764 \\
 & pre\_score & 6.40E-03 & 0.080022  \\
students\_enrolled & (Intercept) & 7.06E-06 & 0.002658 \\
student\_to\_la\_ratio & (Intercept) & 1.53E-03 & 0.039062 \\
pre\_tests\_completed & (Intercept) & 1.54E-01 & 0.392324 \\
post\_tests\_completed & (Intercept) & 1.74E-03 & 0.041762 \\
number\_of\_las & (Intercept) & 2.66E+00 & 1.630497 \\
community\_college & (Intercept) & 4.26E+00 & 2.062855 \\
FCI & (Intercept) & 1.22E+02 & 11.03366 \\
\end{tabular}
\caption{Mixed effects estimates for LASSO study.}
\end{table*}

\begin{table*}[]
\begin{tabular}{ll}
\hline \\
\textbf{Variable}                & \textbf{Feature Importance} \\
\hline \\
pre\_score                       & 462.408209                  \\
is\_imputed\_post\_answered      & 333.693738                  \\
is\_imputed\_post\_score         & 275.880012                  \\
is\_imputed\_log\_post\_duration & 222.376397                  \\
is\_imputed\_log\_pre\_duration  & 153.638948                  \\
FMCE                             & 71.928432                   \\
log\_post\_duration              & 68.916234                   \\
black                            & 68.684196                   \\
hispanic                         & 59.543799                   \\
white                            & 58.978113                   \\
post\_tests\_completed           & 51.276707                   \\
race\_other\_reduced             & 40.774619                   \\
students\_enrolled               & 39.776663                   \\
gender                           & 38.659470                   \\
student\_to\_la\_ratio           & 35.592912                   \\
pre\_tests\_completed            & 28.820861                   \\
number\_of\_las                  & 27.963471                   \\
asian                            & 23.234682                   \\
course\_id                       & 22.554571                   \\
community\_college               & 20.484110                   \\
log\_pre\_duration               & 16.756483                   \\
year\_3rd\_or\_greater           & 15.908271                   \\
year\_2nd                        & 14.254940                   \\
pre\_answered                    & 13.262774                   \\
post\_answered                   & 12.641225                   \\
year\_1st                        & 9.258898                   
\end{tabular}
\caption{Feature importances for gradient boosted model. The feature importance in this table represents the gain \cite{chen2016xgboost}.}
\end{table*}

\clearpage

\subsection{E-CLASS study}

The E-CLASS study presented three separate models: 1)an ordinary least squares (OLS) regression,  2) a mixed effects regression, and 3) a gradient boosted regression.The  feature  coefficients  and  importances  are  presented below.

\begin{table*}[]
\begin{tabular}{lllllll}
\hline \\
\textbf{Variable} & \textbf{coef} & \textbf{std err} & \textbf{z} & $P>|z|$ & \textbf{{[}0.025} & \textbf{0.975{]}} \\
\hline \\
Gender & 0.1637 & 0.074 & 2.224 & 0.026 & 0.019 & 0.308 \\
Q53\_1 & -2.1076 & 0.085 & -24.755 & 0.000 & -2.274 & -1.941 \\
Q53\_2 & 0.0992 & 0.076 & 1.305 & 0.192 & -0.050 & 0.248 \\
Q53\_3 & -0.1475 & 0.079 & -1.870 & 0.062 & -0.302 & 0.007 \\
Q53\_4 & 0.4417 & 0.097 & 4.536 & 0.000 & 0.251 & 0.632 \\
Q53\_5 & 0.0654 & 0.102 & 0.644 & 0.519 & -0.134 & 0.265 \\
Q53\_6 & -0.1277 & 0.141 & -0.908 & 0.364 & -0.403 & 0.148 \\
Q53\_7 & 0.2391 & 0.118 & 2.031 & 0.042 & 0.008 & 0.470 \\
Q53\_8 & 0.2752 & 0.115 & 2.397 & 0.017 & 0.050 & 0.500 \\
Race\_Asian & -0.2432 & 0.100 & -2.441 & 0.015 & -0.438 & -0.048 \\
Race\_Black & -0.6370 & 0.234 & -2.723 & 0.006 & -1.095 & -0.179 \\
Race\_Hispanic/Latino & -1.0918 & 0.186 & -5.865 & 0.000 & -1.457 & -0.727 \\
Race\_Other & -0.2945 & 0.103 & -2.860 & 0.004 & -0.496 & -0.093 \\
Q15\_4 year college & 1.0908 & 0.088 & 12.414 & 0.000 & 0.919 & 1.263 \\
Q15\_Master's granting institution & -0.4627 & 0.148 & -3.120 & 0.002 & -0.753 & -0.172
\end{tabular}
\caption{The logistic regression results for the E-CLASS study.}
\end{table*}

\begin{table*}[]
\begin{tabular}{lll}
\hline \\
Variable & Coefficient & \textbf{Standard Error} \\
\hline \\
Intercept & 2.3514 & 0.0366 \\
Gender & 0.0379 & 0.0214 \\
Q53\_1 & -2.2753 & 0.0210 \\
Q53\_2 & 0.1316 & 0.0220 \\
Q53\_3 & -0.3066 & 0.0216 \\
Q53\_4 & 0.0661 & 0.0195 \\
Q53\_5 & -0.2344 & 0.0196 \\
Q53\_6 & -0.3018 & 0.0189 \\
Q53\_7 & 0.0780 & 0.0196 \\
Q53\_8 & -0.0294 & 0.0192 \\
Race\_Asian & 0.5905 & 0.0903 \\
Race\_Black & 0.2507 & 0.2267 \\
Race\_Hispanic\_Latino & -0.0050 & 0.1702 \\
Race\_Other & 0.5130 & 0.0933 \\
Race\_White & 1.0023 & 0.0469 \\ \\
\textbf{Group} & \textbf{Variance} & \textbf{Standard Deviation} \\
\hline \\
Institution Type (Intercept) & -0.3489 & 0.3983 \\
Institution Type (slope) & -1.8655 & 0.5531
\end{tabular}
\caption{Mixed effects results for E-CLASS study.}
\end{table*}

\begin{table*}[]
\begin{tabular}{ll}
\hline \\
Variable & \textbf{Feature Importance} \\
\hline \\
Q53\_1 & 7.889858 \\
Q53\_3 & 0.735983 \\
Race\_White & 0.505103 \\
Q53\_2 & 0.437148 \\
Q53\_5 & 0.374478 \\
Q15\_4 year college & 0.315693 \\
Q53\_8 & 0.089574 \\
Q15\_Master's granting institution & 0.076935 \\
Q53\_6 & 0.075626 \\
Q53\_4 & 0.071056 \\
Q53\_7 & 0.065396 \\
Race\_Hispanic/Latino & 0.040842 \\
Race\_Asian & 0.038410 \\
Race\_Other & 0.022045 \\
Gender & 0.018761 \\
Race\_Black & 0.005983 \\
Q15\_PhD granting institution & 0.003979
\end{tabular}
\caption{Feature importances for the gradient boosted model for the E-CLASS study. The feature importance in this table represents the gain \cite{chen2016xgboost}.}
\end{table*}

\clearpage
\bibliographystyle{apsrev4-2}
\bibliography{bibliography}

\end{document}